\documentclass[fleqn,twoside]{article}
\usepackage{espcrc2}

\usepackage{graphicx}
\usepackage[figuresright]{rotating}

\newcommand{\AmS}{{\protect\the\textfont2
  A\kern-.1667em\lower.5ex\hbox{M}\kern-.125emS}}

\newcommand{\BR}{{\cal B}}
\newcommand{\psp}{\psi(2S)}
\newcommand{\jpsi}{J/\psi}
\newcommand{\pipi}{\pi^+\pi^-}
\newcommand{\kskp}{K^0_S K^+ \pi^- + c.c.}
\newcommand{\kk}{K^+K^-}
\newcommand{\ppb}{p\overline{p}}

\title{Recent BES Results on Hadron Spectroscopy}

\author{Chang-Zheng Yuan (for the BES Collaboration)\address{Institute of High
Energy Physics, Chinese Academy of Sciences, \\
P.O. Box 918-1, Beijing 100049, China}%
\thanks{Invited talk at the International workshop
on $e^+e^-$ collisions from Phi to Psi, Laboratori Nazionali di
Frascati, Italy, 7-10 April, 2008. Supported in part by the 100
Talents Program of CAS under Contract No.~U-25 and by National
Natural Science Foundation of China under Contract No.~10491303.}}
\begin{document}

\begin{abstract}
We present recent results from the BES experiment on the
observation of the $Y(2175)$ in $J/\psi\to \phi f_0(980) \eta$,
study of $\eta(2225)$ in $J/\psi\to \gamma \phi \phi$, and the
production of $X(1440)$ recoiling against an $\omega$ or a $\phi$
in $J/\psi$ hadronic decays. The observation of $\psi(2S)$
radiative decays is also presented.
 \vspace{1pc}
\end{abstract}

\maketitle

\section{Introduction}

The analyses reported in this talk were performed using either a
sample of $58 \times 10^{6}$ $J/\psi$ events or a sample of $14
\times 10^{6}$ $\psi(2S)$ events collected with the upgraded
Beijing Spectrometer (BESII) detector~\cite{BESII} at the Beijing
Electron-Positron Collider (BEPC).

\section{The $Y(2175)$ in $J/\psi\to \phi f_0(980) \eta$~\cite{bes2175}}

A new structure, denoted as $Y(2175)$ and with mass $m=2.175\pm
0.010\pm 0.015$~GeV/$c^2$ and width $\Gamma=58\pm 16\pm
20$~MeV/$c^2$, was observed by the BaBar experiment in the
$e^+e^-\to\gamma_{ISR}\phi f_0(980)$ initial-state radiation
process~\cite{babary21752006,babary21752007}. This observation
stimulated some theoretical speculation that this $J^{PC}=1^{--}$
state may be an $s$-quark version of the $Y(4260)$ since both of
them are produced in $e^+e^-$ annihilation and exhibit similar
decay patterns~\cite{babar4260,belle4260}.

\begin{figure}[htbp]
  \centering
\includegraphics[width=0.40\textwidth]{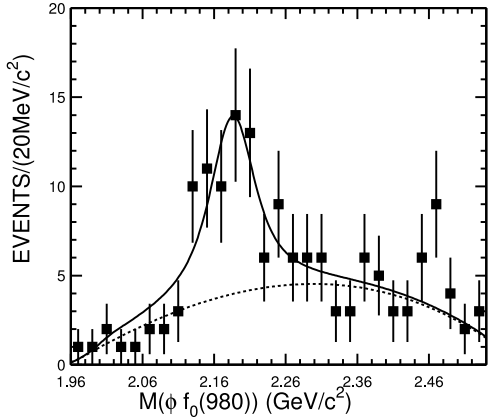}
\caption{The $\phi f_0(980)$ invariant mass distribution of the
data (points with error bars) and the fit (solid curve) with a
Breit-Wigner function and polynomial background; the dashed curve
indicates the background function.} \label{draft-fit}
\end{figure}

Here we report the observation of the $Y(2175)$ in the decays of
$J/\psi\to \eta \phi f_0(980)$, with $\eta\to \gamma\gamma$, $\phi
\to K^+K^-$, $f_0(980)\to\pi^+\pi^-$. A four-constraint
energy-momentum conservation kinematic fit is performed to the
$K^+ K^-\pi^+\pi^-\gamma\gamma$ hypothesis for the selected four
charged tracks and two photons. $\eta\to\gamma\gamma$ candidates
are defined as $\gamma$-pairs with
$|M_{\gamma\gamma}-0.547|<0.037$~GeV/$c^2$, a $\phi$ signal is
defined as $|m_{K^+K^-}-1.02|<0.019$~GeV/$c^2$, and in the
$\pi^+\pi^-$ invariant mass spectrum, candidate $f_0(980)$ mesons
are defined by $|m_{\pi^+\pi^-}-0.980|<0.060$~GeV/$c^2$. The $\phi
f_0(980)$ invariant mass spectrum for the selected events is shown
in Fig.~\ref{draft-fit}, where a clear enhancement is seen around
2.18~GeV/$c^2$. Fit with a Breit-Wigner and a polynomial
background yields $52\pm12$ signal events and the statistical
significance is found to be $5.5\sigma$ for the signal. The mass
of the structure is determined to be $M=2.186\pm 0.010~(stat)\pm
0.006~(syst)$~GeV/$c^2$, the width is $\Gamma=0.065\pm
0.023~(stat)\pm 0.017~(syst)$~GeV/$c^2$, and the product branching
ratio is $\BR(J/\psi \to \eta Y(2175))\cdot \BR(Y(2175)\to \phi
f_0(980))\cdot \BR(f_0(980)\to\pi^+\pi^-)=(3.23\pm 0.75~(stat)\pm
0.73~(syst))\times 10^{-4}$. The mass and width are consistent
with BaBar's results.

\section{The $\eta(2225)$ in $J/\psi\to \gamma \phi\phi$~\cite{bes2225}}

Structures in the $\phi\phi$ invariant-mass spectrum have been
observed by several experiments both in the reaction
$\pi^{-}p\to\phi\phi n$~\cite{pip} and in radiative $J/\psi$
decays~\cite{mk3,dm21,dm22}. The $\eta(2225)$ was first observed
by the MARK-III collaboration in $J/\psi$ radiative decays $J/\psi
\to \gamma\phi\phi$. A fit to the  $\phi\phi$ invariant-mass
spectrum gave a mass of 2.22~GeV/$c^2$ and a width of
150~MeV/$c^2$~\cite{mk3}. An angular analysis of the structure
found it to be consistent with a $0^{-+}$ assignment. It was
subsequently observed by the DM2 collaboration, also in $J/\psi
\to \gamma \phi \phi$ decays~\cite{dm21,dm22}.

\begin{figure}[htbp]
  \centering
\includegraphics[width=0.40\textwidth]{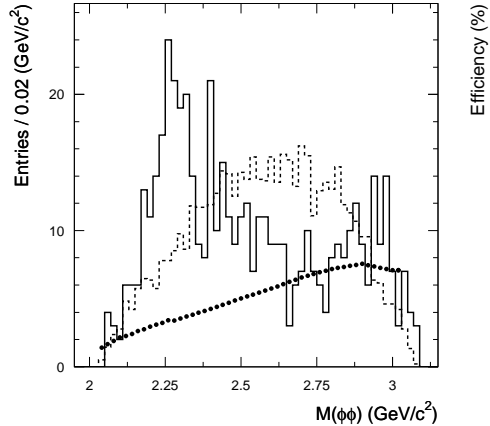}
\caption{The $K^+K^-K^0_SK^0_L$ invariant mass distribution for
$J/\psi \to \gamma \phi\phi$ candidate events. The dashed
histogram is the phase space invariant mass distribution, and the
dotted curve indicates how the acceptance varies with the
$\phi\phi$ invariant mass.}   \label{dalitz}
\end{figure}

We present results from a high statistics study of $J/\psi \to
\gamma \phi \phi$ in the $\gamma K^+ K^- K^0_S K^0_L$ final state,
with the $K^0_L$ missing and reconstructed with a one-constraint
kinematic fit. After kinematic fit, we require both the $K^+K^-$
and $K^0_SK^0_L$ invariant masses lie within the $\phi$ mass
region ($|M(K^+K^-)-m_{\phi}|<12.5$~MeV/$c^2$ and
$|M(K^0_SK^0_L)-m_{\phi}|<25$~MeV/$c^2$). The $\phi\phi$ invariant
mass distribution is shown in Fig.~\ref{dalitz}. There are a total
of 508 events with a prominent structure around 2.24~GeV/$c^2$.

A partial wave analysis of the events with $M(\phi\phi)<$
2.7~GeV/$c^2$ was performed. The two-body decay amplitudes in the
sequential decay process $J/\psi \to \gamma X, X\to \phi\phi$,
$\phi\to K^+ K^- $ and $\phi\to K^0_S K^0_L$ are constructed using
the covariant helicity coupling amplitude method. The intermediate
resonance $X$ is described with the normal Breit-Wigner propagator
$BW = 1/(M^2-s-i M\Gamma)$, where $s$ is the $\phi\phi$ invariant
mass-squared and $M$ and $\Gamma$ are the resonance's mass and
width. When $J/\psi \to \gamma X$, $X \to \phi \phi$ is fitted
with both the $\phi\phi$ and $\gamma X$ systems in a $P$-wave,
which corresponds to a pseudoscalar $X$ state, the fit gives
$196\pm 19$ events with mass $M
=2.24^{+0.03}_{-0.02}{}^{+0.03}_{-0.02}$~GeV/$c^2$, width $\Gamma
=0.19\pm0.03^{+0.04}_{-0.06}$~GeV/$c^2$, and a statistical
significance larger than $10\sigma$, and a product branching
fraction of: $\BR(J/\psi \to \gamma \eta(2225))\cdot
\BR(\eta(2225)\to \phi\phi)=(4.4\pm 0.4\pm 0.8)\times 10^{-4}$.

The presence of a signal around 2.24~GeV/$c^2$ and its
pseudoscalar character are confirmed, and the mass, width, and
branching fraction are in good agreement with previous
experiments.

\section{The $X(1440)$ in $J/\psi$ hadronic decays~\cite{bes1440}}

A pseudoscalar gluonium candidate, the so-called $E/\iota(1440)$,
was observed in $p\bar{p}$ annihilation in 1967~\cite{baillon67}
and in $J/\psi$ radiative decays in the
1980's~\cite{scharre80,edwards82e,augustin90}. The study of the
decays $J/\psi \rightarrow$ \{$\omega$, $\phi$\}$K\bar{K}\pi$ is a
useful tool in the investigation of quark and possible gluonium
content of the states around 1.44~GeV/$c^{2}$.  Here we
investigate the possible structure in the $K\bar{K}\pi$ final
state in $J/\psi$ hadronic decays at around $1.44$~GeV/$c^{2}$.

In this analysis, $\omega$ mesons are observed in the $\omega
\rightarrow \pi^{+}\pi^{-}\pi^{0}$ decay, $\phi$ mesons in the
$\phi \rightarrow K^{+}K^{-}$ decay, and other mesons are detected
in the decays: $K^{0}_{S} \rightarrow \pi^{+}\pi^{-}$, $\pi^0
\rightarrow \gamma \gamma$. $K\bar{K}\pi$ could be
$K^{0}_{S}K^{\pm} \pi^{\mp}$ or $K^{+}K^{-}\pi^{0}$.

\begin{figure*}[htbp]
  \centering
\includegraphics[width=0.43\textwidth]{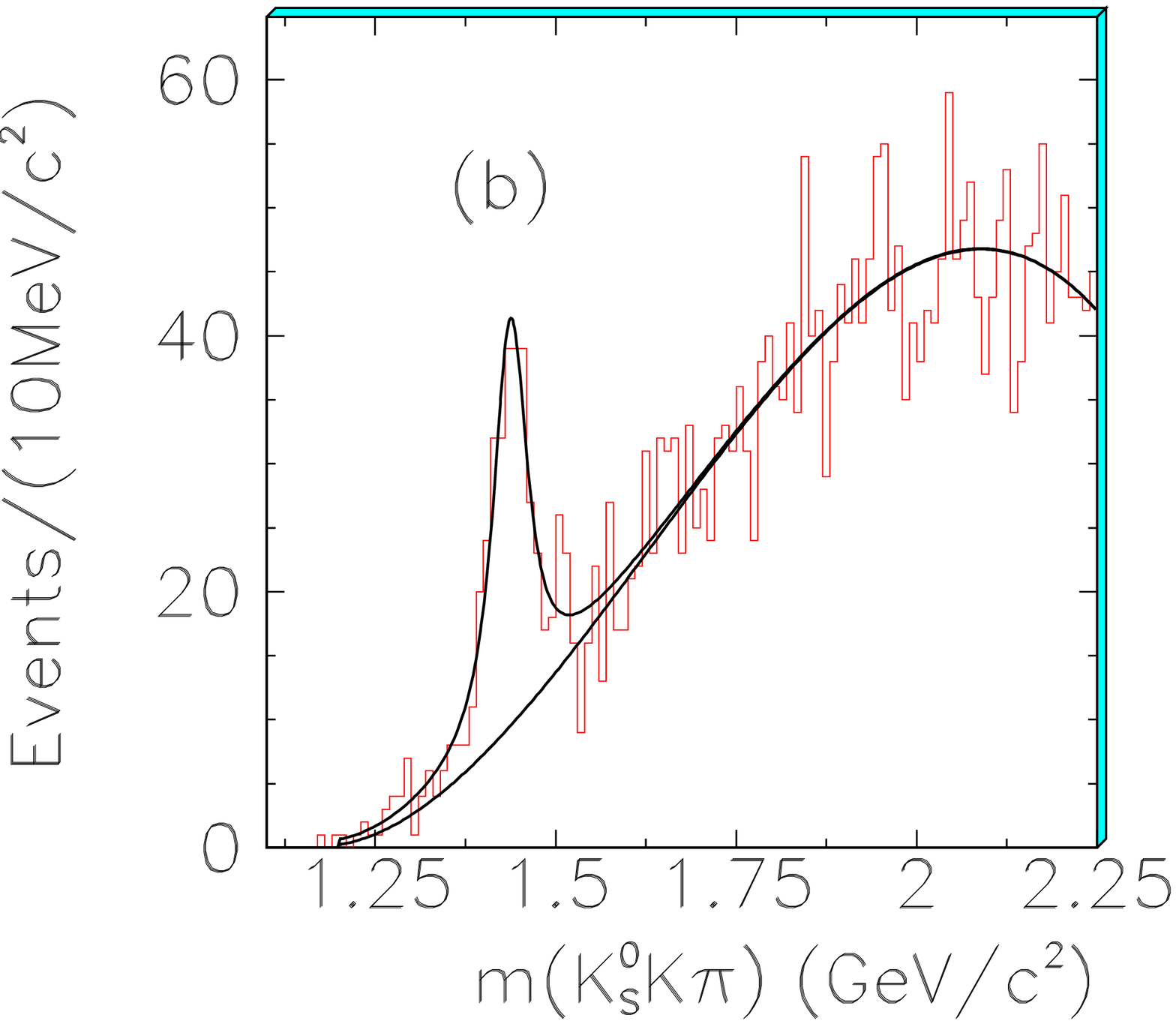}
\includegraphics[width=0.43\textwidth]{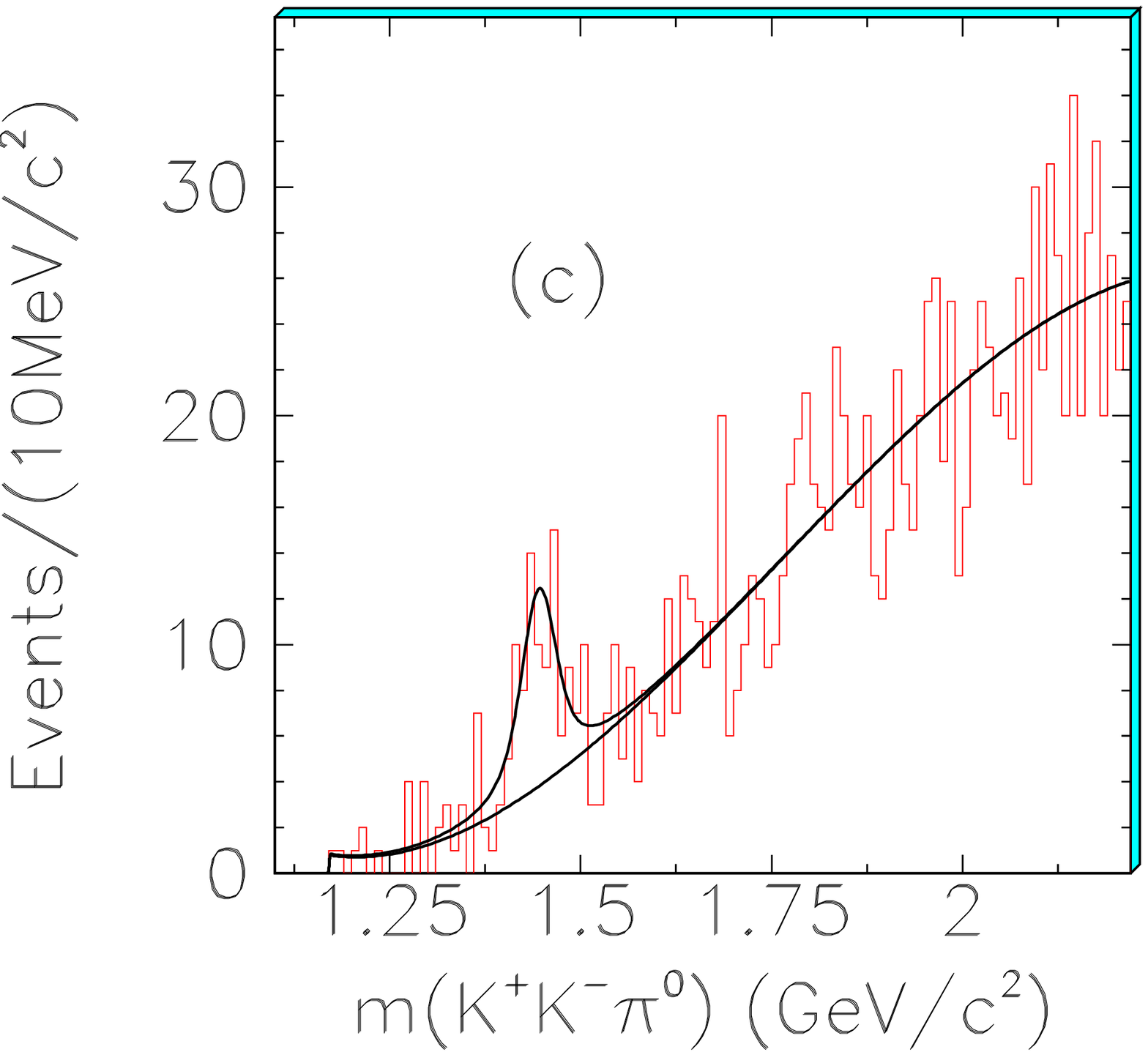}
\caption{The $K\bar{K}\pi$ invariant mass distribution for $J/\psi
\rightarrow \omega K^{0}_{S}K^{\pm}\pi^{\mp}$ (b) and $\omega
K^{+}K^{-}\pi^{0}$ (c) candidate events.  The curves are the best
fit. } \label{fig:w-x1440-recoiling}
\end{figure*}

\begin{figure*}[htbp]
  \centering
\includegraphics[width=0.43\textwidth]{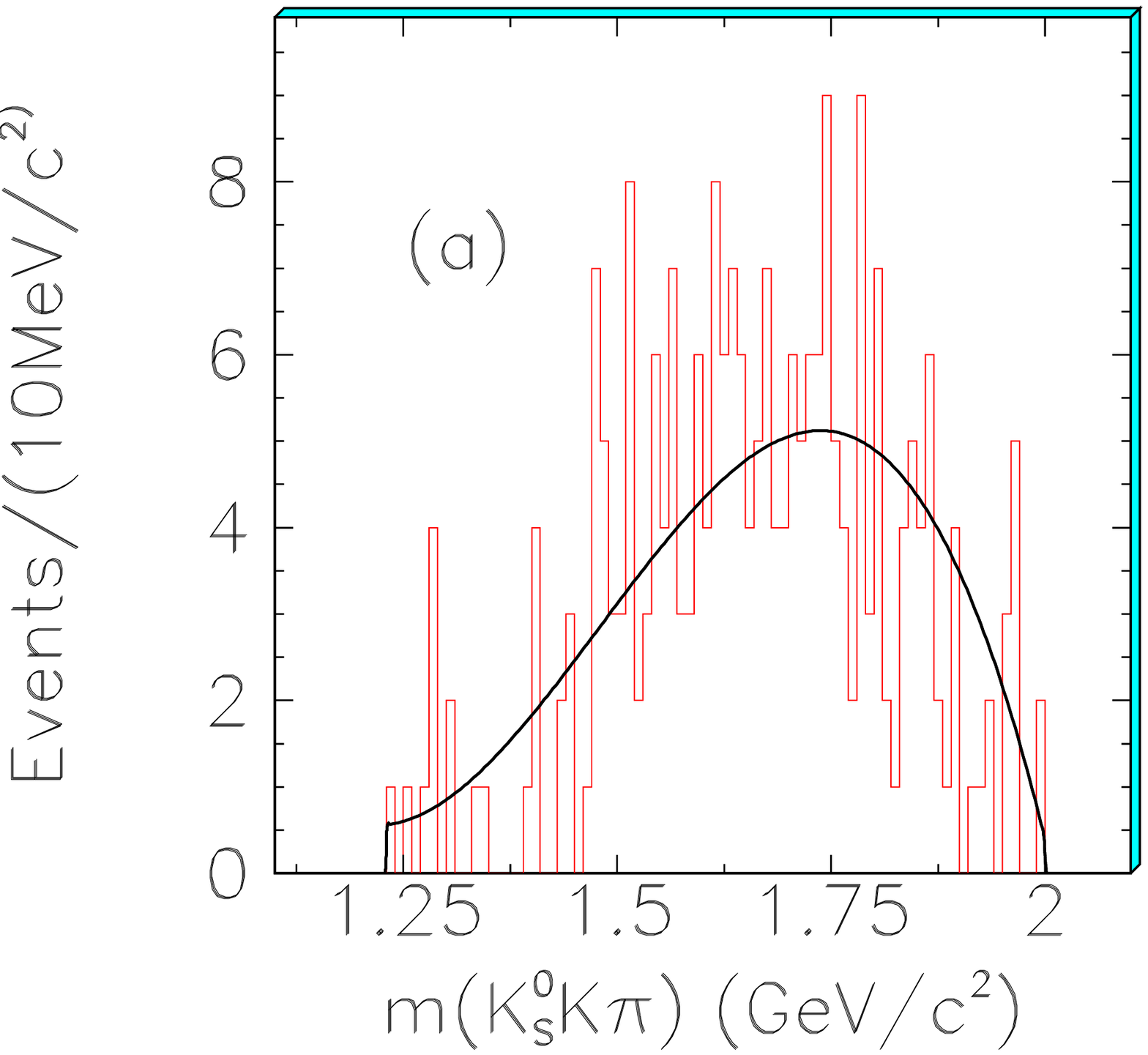}
\includegraphics[width=0.43\textwidth]{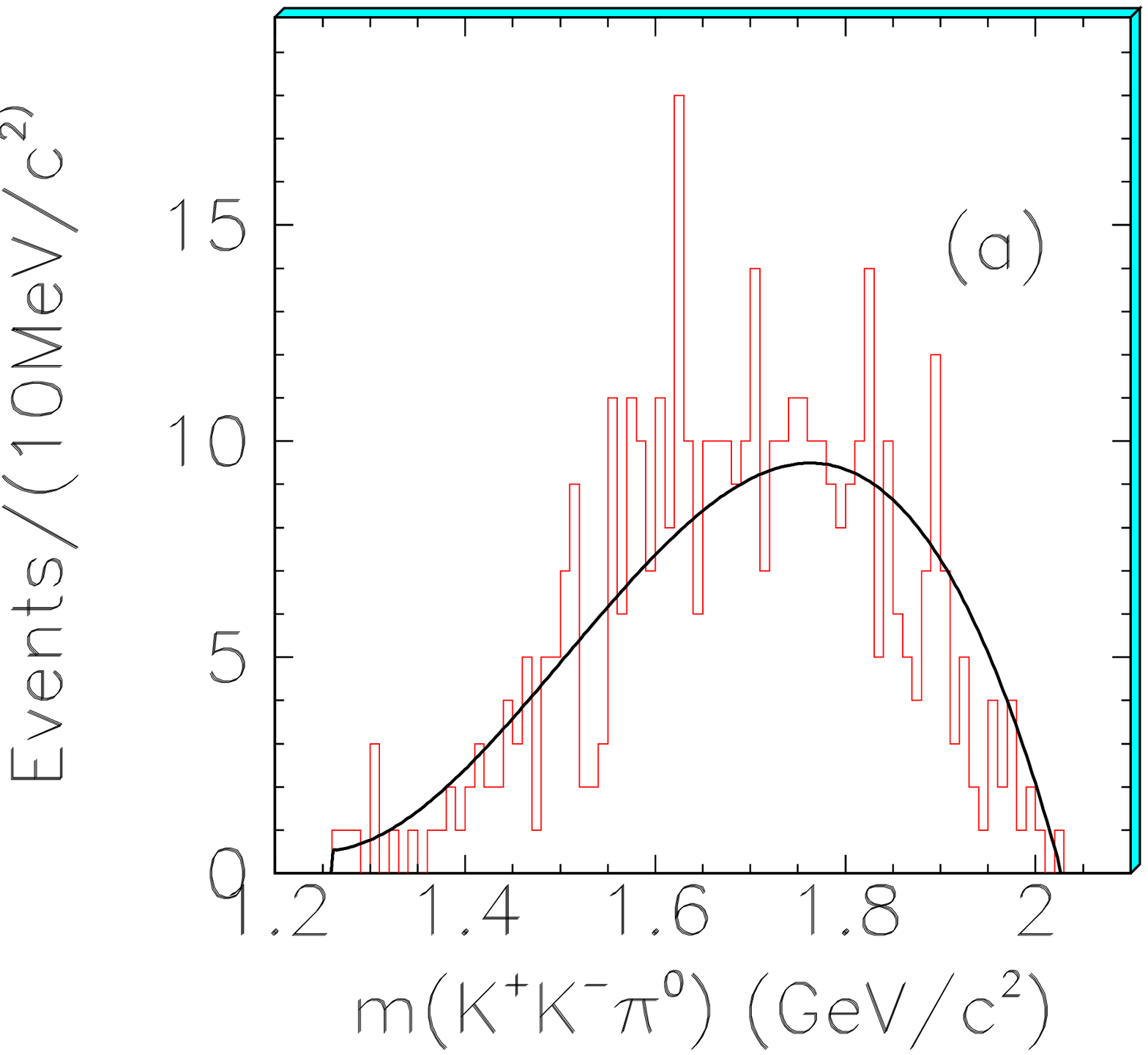}
\caption{The $K^{0}_{S} K^{\pm} \pi^{\mp}$  (left) and $\kk\pi^0$
(right) invariant mass recoiling against the $\phi$ in $\jpsi\to
\phi K\bar{K}\pi$ mode.} \label{fig:x1440-phikksp}
\end{figure*}

Figures~\ref{fig:w-x1440-recoiling} and~\ref{fig:x1440-phikksp}
show the $K^{0}_{S}K^{\pm}\pi^{\mp}$ and $K^{+} K^{-}\pi^{0}$
invariant mass spectra after $\omega$ selection
($|m_{\pi^{+}\pi^{-}\gamma \gamma}-m_{\omega}|<0.04$ GeV/c$^{2}$)
or $\phi$ signal selection
($|m_{K^{+}K^{-}}-m_{\phi}|<0.015$~GeV/$c^{2}$). Clear $X(1440)$
signal is observed recoiling against the $\omega$, and there is no
significant signal recoiling against a $\phi$.

The $K^{0}_{S}K^{\pm}\pi^{\mp}$ invariant mass distribution in
$\jpsi\to \omega K^{0}_{S}K^{\pm}\pi^{\mp}$
(Fig.~\ref{fig:w-x1440-recoiling}(b)) is fitted with a BW function
convoluted with a Gaussian mass resolution function
($\sigma=7.44$~MeV/$c^{2}$) to represent the $X(1440)$ signal and
a third-order polynomial background function. The mass and width
obtained from the fit are $M=1437.6\pm 3.2$~MeV/$c^{2}$ and
$\Gamma=48.9 \pm 9.0$~MeV/$c^{2}$, and the fit yields $249\pm 35$
events. Using the efficiency of $1.45\%$ determined from a uniform
phase space MC simulation, we obtain the branching fraction to be
$\BR(J/\psi \rightarrow \omega X(1440))\cdot \BR( X(1440)
\rightarrow K^{0}_{S} K^{+} \pi^{-}+c.c.) = (4.86\pm 0.69\pm 0.81)
\times 10^{-4}$, where the first error is statistical and the
second one systematic.

For $\jpsi\to \omega K^{+} K^{-}\pi^{0}$ mode, by fitting the
$K^{+} K^{-}\pi^{0}$ mass spectrum in
Fig.~\ref{fig:w-x1440-recoiling}(c) with same functions, we obtain
the mass and width of $M=1445.9\pm 5.7$~MeV/$c^{2}$ and
$\Gamma=34.2\pm 18.5$~MeV/$c^{2}$, and the number of events from
the fit is $62\pm 18$. The efficiency is determined to be $0.64\%$
from a phase space MC simulation, and the branching fraction is
$\BR(J/\psi \rightarrow \omega X(1440)) \cdot \BR(X(1440)
\rightarrow K^{+} K^{-} \pi^{0}) = (1.92\pm 0.57\pm 0.38) \times
10^{-4}$, in good agreement with the isospin symmetry expectation
from $\jpsi\to \omega K^{0}_{S}K^{\pm}\pi^{\mp}$ mode.

The distribution of $K^{0}_{S} K^{\pm} \pi^{\mp}$ and $K^{+} K^{-}
\pi^{0}$ invariant mass spectra recoiling against the $\phi$
signal are shown in Fig.~\ref{fig:x1440-phikksp}, and there is no
evidence for $X(1440)$. The upper limits on the branching
fractions at the $90\%$ C.L. are $\BR(J/\psi\rightarrow \phi
X(1440) \rightarrow \phi K^{0}_{S}K^{+}\pi^{-}+c.c.)<1.93 \times
10^{-5}$ and $\BR( J/\psi \rightarrow \phi X(1440) \rightarrow
\phi K^{+}K^{-}\pi^{0}) < 1.71 \times 10^{-5}$.

In conclusion, the mass and width of the $X(1440)$ are measured,
which are in agreement with previous measurements; the branching
fractions we measured are also in agreement with the DM2 and
MARK-III results. The significant signal in $\jpsi\to \omega
K\bar{K}\pi$ mode and the missing signal in $\jpsi\to \phi X$ mode
may indicate the $s\bar{s}$ component in the $X(1440)$ is not
significant.

\section{\boldmath $\psp$ radiative decays}

Besides conventional meson and baryon states, QCD also predicts a
rich spectrum of glueballs, hybrids, and multi-quark states in the
1.0 to 2.5~$\hbox{GeV}/c^2$ mass region. Therefore, searches for
the evidence of these exotic states play an important role in
testing QCD. The radiative decays of $\psp$ to hadrons are
expected to contribute about 1\% to the total $\psp$ decay
width~\cite{PRD_wangp}. However, the measured channels only sum up
to about 0.05\%~\cite{PDG}.

\begin{figure}
  \centering
\includegraphics[width=0.48\textwidth,height=0.45\textheight]{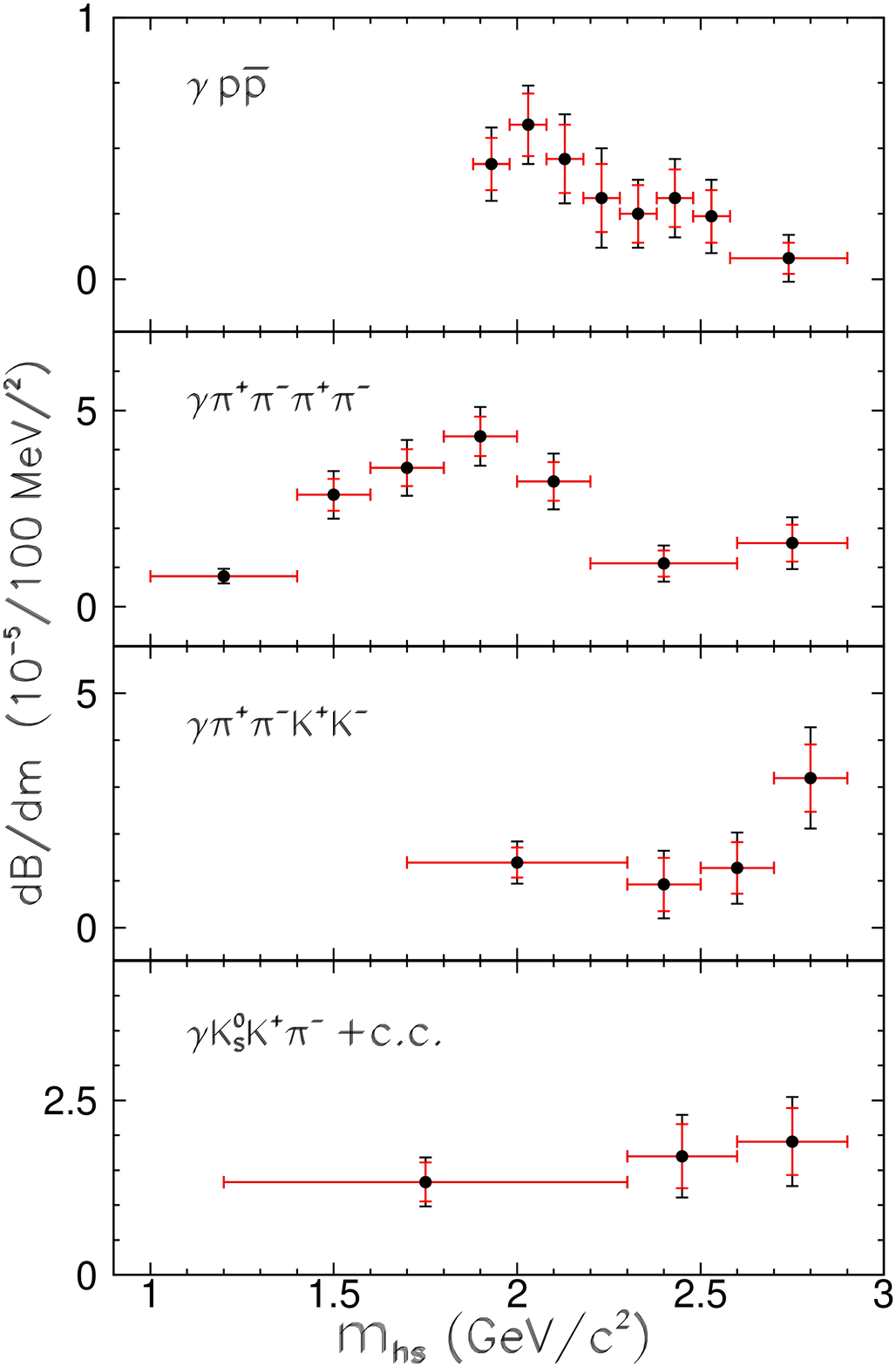}
\caption{ \label{difbr} Differential branching fractions for
$\psp\to \gamma\ppb$, $\gamma 2(\pipi)$, $\gamma K^+ K^- \pipi$,
and $\gamma \kskp$. Here $m_{hs}$ is the invariant mass of the
hadrons. For each point, the smaller longitudinal error is the
statistical error, while the bigger one is the total error. }
\end{figure}

We measured the decays of $\psp$ into $\gamma\ppb$, $\gamma
2(\pipi)$, $\gamma \kskp$, $\gamma K^+ K^- \pipi$, $\gamma
K^{*0}K^-\pi^+ +c.c.$, $\gamma K^{*0}\bar K^{*0}$,
$\gamma\pipi\ppb$, $\gamma 2(\kk)$, $\gamma 3(\pipi)$, and $\gamma
2(\pi^+\pi^-)K^+K^-$, with the invariant mass of the hadrons
($m_{hs}$) less than 2.9~$\hbox{GeV}/c^2$ for each decay
mode~\cite{bes2rad}. The differential branching fractions are
shown in Fig.~\ref{difbr}. The branching fractions below
$m_{hs}<2.9$~$\textrm{GeV}/c^2$ are given in Table~\ref{Tot-nev},
which sum up to $0.26\%$ of the total $\psp$ decay width. We also
analyzed $\psp\to \gamma\pipi$ and $\gamma\kk$ modes to study the
resonances in $\pipi$ and $\kk$ invariant mass spectrum.
Significant signals for $f_2(1270)$ and $f_0(1710)$ were observed,
but the low statistics prevent us from drawing solid conclusion on
the other resonances~\cite{agnes}.

\begin{table}
\caption{\label{Tot-nev} Branching fractions for $\psp\to\gamma
+hadrons$ with $m_{hs}<2.9$ $\hbox{GeV}/c^2$, where the upper
limits are determined at the 90\% C.L.}
\begin{center}
\begin{tabular}{ll} \hline \hline
Mode & $\BR(\times 10^{-5})$\\\hline
$\gamma p\bar{p}$ & 2.9$\pm$0.4$\pm$0.4 \\
$\gamma 2(\pi^+\pi^-)$ & 39.6$\pm$2.8$\pm$5.0\\
$\gamma K^0_S K^+\pi^-+c.c.$  & 25.6$\pm$3.6$\pm$3.6 \\
$\gamma K^+ K^-\pi^+\pi^-$ & 19.1$\pm$2.7$\pm$4.3 \\
$\gamma K^{*0} K^+\pi^-+c.c.$& 37.0$\pm$6.1$\pm$7.2\\
$\gamma K^{*0}\bar K^{*0}$&$24.0\pm 4.5\pm 5.0$\\
$\gamma \pipi\ppb$& 2.8$\pm$1.2$\pm$0.7 \\
$\gamma \kk\kk$ &  $<4$\\
$\gamma3(\pipi)$&  $<17$\\
$\gamma2(\pi^+\pi^-)K^+K^-$& $<22$ \\
\hline \hline
\end {tabular}
\end{center}
\end{table}

\section{\boldmath Summary}

Using the 58~M $\jpsi$ and 14~M $\psp$ events samples taken with
the BESII detector at the BEPC storage ring, BES experiment
provided many interesting results in charmonium decays, including
the observation of the $Y(2175)$, $\eta(2225)$, $X(1440)$, and
many $\psp$ radiative decays. These results shed light on the
understanding of strong interaction sector of the Standard Model.

\end{document}